\documentclass[%
 aps,prb,superscriptaddress,twocolumn
]{revtex4-2}

\DeclareUnicodeCharacter{2225}{\ensuremath{\parallel}}

\usepackage{graphicx}
\usepackage{dcolumn}
\usepackage{bm}
\usepackage{color}
\usepackage{comment}
\usepackage{mathrsfs}
\usepackage{amsmath,amssymb}

\begin{document}

\preprint{APS/123-QED}

\title{\textbf{Large magnetostriction and first-order phase transition\\ induced by ultrahigh magnetic fields in V$_6$O$_{13}$} 
}%

\author{Y.Ishii}
\affiliation{Institute for Solid State Physics, University of Tokyo, Kashiwa, Chiba 277-8581, Japan}

\author{A.Ikeda}
\affiliation{Department of Engineering Science, University of Electro-Communications, Chofu, Tokyo, 182-8585, Japan}

\author{Y.H.Matsuda}
\affiliation{Institute for Solid State Physics, University of Tokyo, Kashiwa, Chiba 277-8581, Japan}


\newcommand{\YI}[1]{{\color[RGB]{0,0,255}{{#1}\,}}}
\newcommand{\AI}[1]{{\color[RGB]{0,255,0}{{#1}\,}}}
\newcommand{\Yhmc}[1]{{\color[RGB]{255,0,0}{{#1}\,}}}

\clearpage

\begin{abstract}
V$_6$O$_{13}$ exhibits the metal-insulator transition (MIT) at around 150 K with the formation of vanadium-vanadium (V-V) dimers.
We measured the magnetostriction of V$_6$O$_{13}$ along the crystallographic $b$-axis in ultrahigh magnetic fields up to 186 T at various temperatures.
An abrupt and large negative magnetostriction of $\Delta L / L \sim 10^{-3}$ was observed above 120 T below 150 K which is attributed to the collapse of the V-V dimer.
A pronounced hysteresis exceeding 50 T was observed, indicating a first-order phase transition.
The field-induced collapse of the V-V dimer is expected to be accompanied by the insulator-to-metal transition.
The robustness of the insulating phase against magnetic fields was discussed based on thermodynamics and the comparison with the case of VO$_2$.
\end{abstract}

\maketitle

\section{Introduction}
The formation of an electronic multimer is one of the most intriguing mechanisms of electronic phase transitions, as it can trigger drastic phase transitions and give rise to exotic electronic states \cite{ZHiroi2015}.
In particular, the metal-insulator transition (MIT) is a representative example of a phase transition associated with electronic multimer formation.

VO$_2$ is a representative MIT material that exhibits vanadium-vanadium (V-V) dimers \cite{FJMorin1959}.
The formation of V-V dimers corresponding to the molecular orbital (MO) is considered one of the dominant mechanisms of MIT \cite{JBGoodenough1971,RJOMossanek2005,KLi2023,ZHiroi2015}.
The $d$ orbitals of neighboring V ions overlap and form molecular orbital states, as revealed by recent detailed X-ray diffraction experiments \cite{SKitou2024_VO2}.
On the other hand, electronic correlation effects \cite{AZylbersztejn1975, DPaquet1980, RMWentzcovitch1994} and Peierls instabilities \cite{SBiermann2005,MGupta1977,VEyert2002} have also been pointed out as an important factor in the MIT of VO$_2$.
The mechanism of MIT in VO$_2$ remains a subject of intense debate.

\begin{figure*}
\includegraphics[clip,width=1.9\columnwidth]{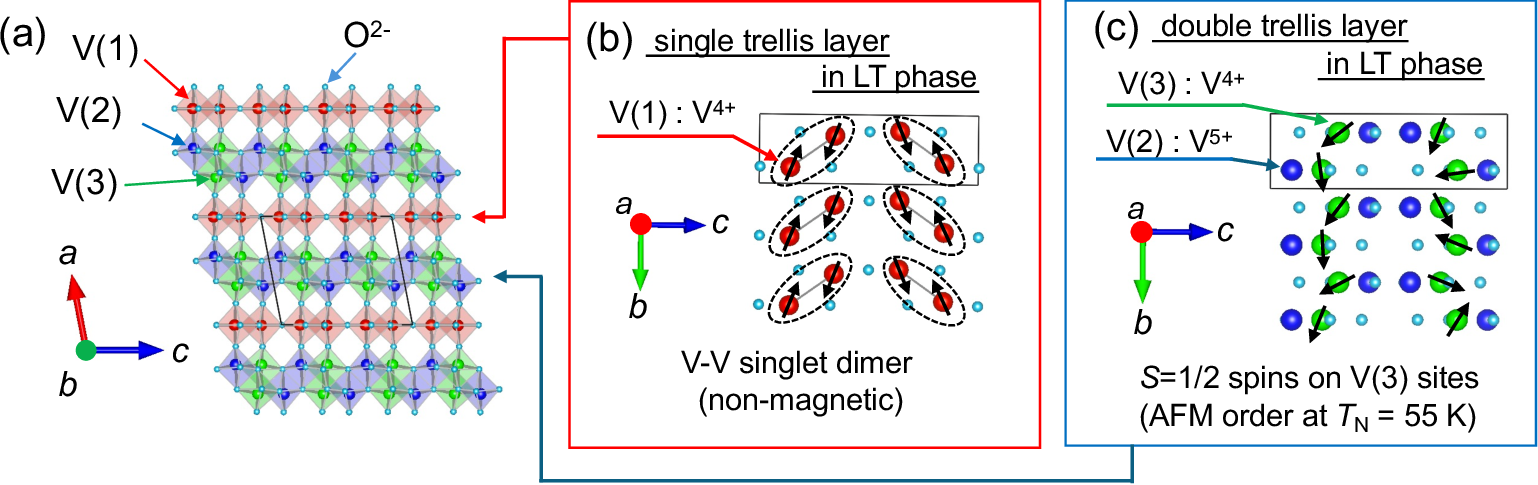}
\caption{\label{image} (Color online) (a) Crystal structure of V$_6$O$_{13}$ along $b$-axis. (b), (c) Schematic image for the electronic and magnetic state in the low temperature insulating phase.}
\end{figure*}

There are several other vanadium oxides that exhibit an MIT.
Among them, V$_6$O$_{13}$ belongs to the Wadsley-phase series V$_{\rm n}$O$_{2{\rm n}+1}$ and is known to exhibit an MIT involving V-V dimers.
V$_6$O$_{13}$ has also been studied as a potential cathode in a Li-ion battery \cite{KWest1985,MYSaidi1997,TSchmitt2005}.
The crystal structure consists of two-dimensional single-trellis and double-trellis layers with monoclinic symmetry, as shown in Fig. \ref{image}(a) \cite{PDDernier1974,Hakala2022}.
Electronic properties have been intensively investigated by photoemission and NMR experiments \cite{SShin1990,REguchi2002,TSchmitt2004,SSuga2004,SSuga2010,YShimizu2015}.
The average valence state of the V ions is 4.33, corresponding to a 1:2 ratio of V$^{5+}$ (3$d^0$) and V$^{4+}$ (3$d^1$) ions.
The single-trellis layer is composed of V(1) ions and the double-trellis layer consists of V(2) and V(3) ions (Fig. \ref{image}(b)).
The V(1), V(2), and V(3) sites are occupied by V$^{4+}$, V$^{5+}$, whereas V$^{4+}$ ions, respectively \cite{PDDernier1974,ACGossard1974}.
The MIT, accompanied by a structural phase transition, occurs at $T_{\rm MIT}$ = 160 K \cite{KKawashima1974}.
At $T_{\rm MIT}$ the V(1) ions in the single-trellis layer form V-V dimers, entering nonmagnetic spin-singlet states (Fig. \ref{image}(b)) \cite{PDDernier1974,Howing2003}.
In contrast, within the double-trellis layer, V(2) corresponds to a nonmagnetic V$^{5+}$ ion, whereas V(3) is a magnetic V$^{4+}$ ion carrying a localized spin of $S = 1/2$ (Fig. \ref{image}(c)).
These detailed electronic and magnetic properties of V$_6$O$_{13}$ have also been clarified by recent theoretical calculations and NMR studies \cite{TToriyama2014,YShimizu2015,IVLeonov2022}.
Furthermore, according to NMR study, the MIT can be understood as a rare case of a site-selective Mott transition (on the V(1) site) combined with an orbitally assisted spin-Peierls transition \cite{YShimizu2015}.
This coexistence of Mott character and dimerization is a particularly intriguing feature and is distinct from the case of VO$_2$.

The ability to control electronic multimers by external fields is crucial as it not only opens avenues for exploring new functionalities of materials but also deepens our understanding of their electronic states.
VO$_2$ shows the pressure-induced structural phase transition \cite{CNBerglund1969,YChen2017} and the current-induced nonequilibrium phase transition \cite{ANakano2025}, providing the new aspect of the material.
The magnetic field effect is also significant because it can selectively probe the roles of spin and orbital degrees of freedom, providing valuable insights into the nature of the electronic state and the underlying mechanisms of the phase transition.

However, to date, the experimental demonstration of magnetic field-induced collapse of V-V dimers has been reported only in VO$_2$ \cite{YHMatsuda2020,YHMatsuda2022}.
It is accompanied by an insulator-metal transition, as evidenced by measurements of optical transmission up to 600 T \cite{YHMatsuda2020,YHMatsuda2022}.
This provides experimental proof that the V-V dimers within solids can collapse under sufficiently strong magnetic fields.
Furthermore, these results also experimentally demonstrate that spin degrees of freedom play an essential role in MIT in VO$_2$ \cite{YHMatsuda2020,YHMatsuda2022}.

Investigating the magnetic field effects on the MIT in V$_6$O$_{13}$ should provide a new perspective on the nature of MIT from the viewpoint of field-induced phenomena and the new experimental demonstration of the field-induced collapse of V-V dimer.
Furthermore, comparison with the case of VO$_2$ will allow us to discuss the magnetic field effect of V-V dimers perspectively.

In this paper, we investigate the effects of magnetic fields on the MIT associated with V-V dimerization in V$_6$O$_{13}$, using magnetostriction measurements up to 186 T.
A pronounced magnetostriction was observed above 120 T below $T_{\rm MIT}$, which is attributed to the field-induced collapse of the V-V dimer.
The large hysteresis indicates that the observed magnetostriction corresponds to a first-order phase transition, which is expected to be accompanied by the insulator-metal transition.
Furthermore, we discuss the relationship between the V-V dimer and the MIT in V$_6$O$_{13}$, the origin of the large hysteresis, the phase boundary based on thermodynamics, and a comparison between V$_6$O$_{13}$ and VO$_2$.

\begin{figure}
\includegraphics[clip,width=0.95\columnwidth]{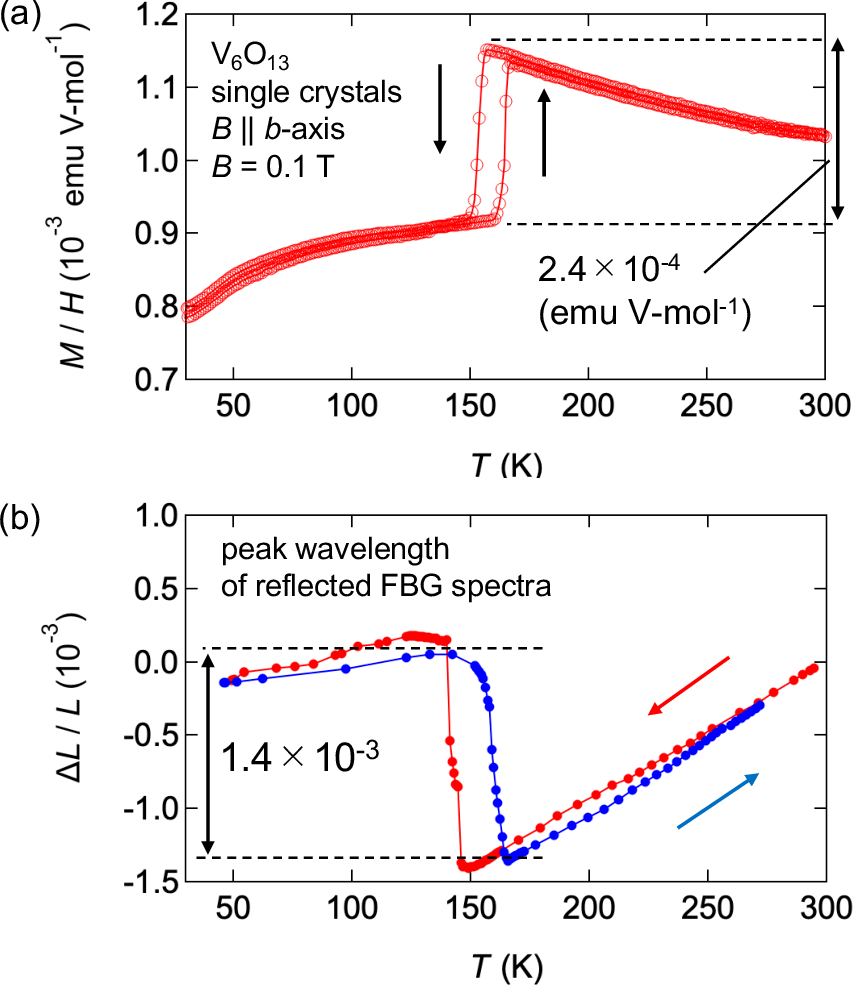}
\caption{\label{ChiT_FBG} (Color online) (a) Temperature dependence of magnetic susceptibility of V$_6$O$_{13}$ single crystals at $B$ = 0.1 T. (b) Temperature dependence of $\Delta L / L$ of V$_6$O$_{13}$ single crystal along the $b$-axis measured by the FBG and optical filter method.}
\end{figure}

\begin{figure*}
\includegraphics[clip,width=1.9\columnwidth]{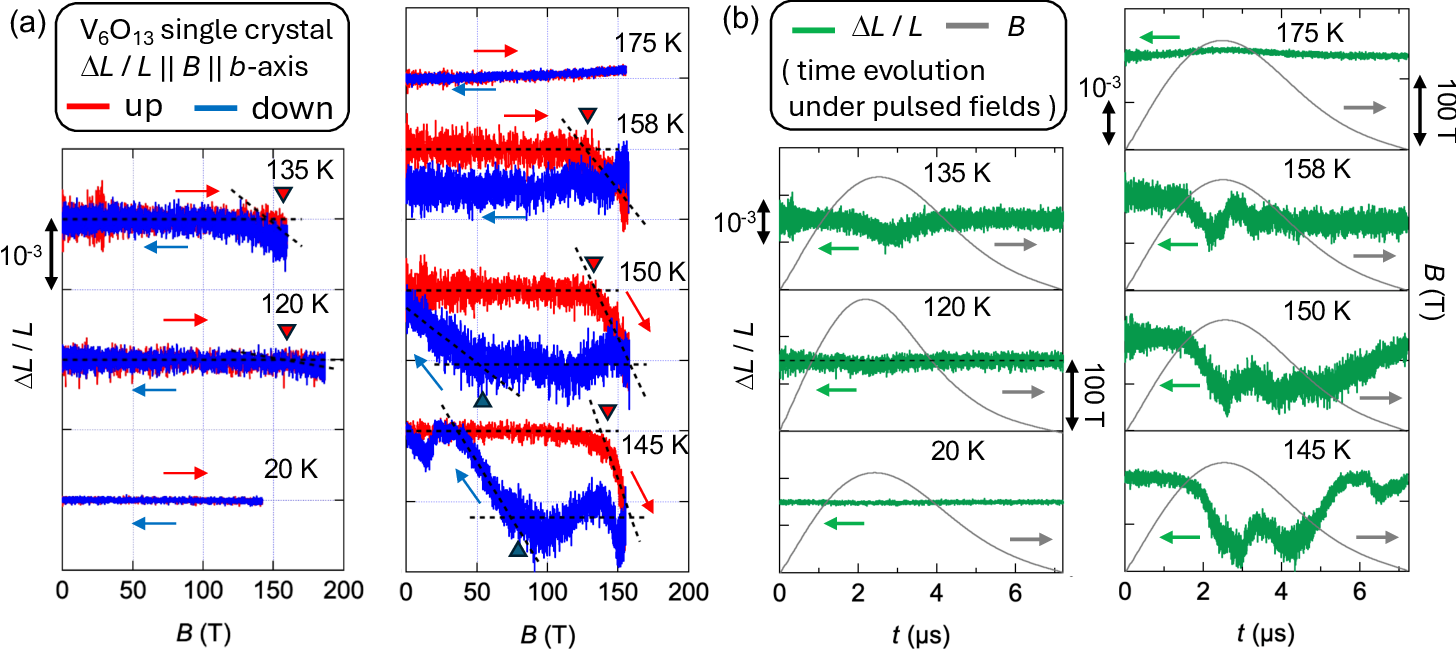}
\caption{\label{FBG_varT} (Color online) (a) Magnetostriction in ultrahigh magnetic fields at several temperatures. Red and blue curves represent the data in the field ascending and descending processes, respectively. Black broken lines are the linear fitting results of each data. (b) Magnetostriction and magnetic field as a function of time at several temperatures. Green and gray curves are the magnetostriction and magnetic field, respectively.}
\end{figure*}

\section{Experiments}
Single crystals of V$_6$O$_{13}$ were synthesized by the chemical vapor transport method reported in Ref. \cite{KNagasawa1969} (a detailed information is summarized in supplementary material \cite{Suppl}).
Magnetic susceptibility was measured by the Magnetic Property Measurement System (Quantum Design, MPMS) on several tens of small single crystals under the condition that the magnetic field is parallel to the $b$-axis.
Pulsed ultrahigh magnetic fields up to 186 T were generated by the horizontal-type single-turn coil (HSTC) system in the Institute for Solid State Physics, the University of Tokyo, Japan \cite{NMiura2003}. 
For temperature control, helium flow-type cryostats made of (nonmetallic) fiber-reinforced plastics were employed to avoid generating an induction current with a high sweep rate of the magnetic field generated by HSTC.
The chromel-constantan thermocouple was used for the sample temperature monitor.
$\Delta L / L$ was measured using the high-speed strain gauge using the fiber Bragg grating (FBG) and the optical filter method \cite{AIkeda2017}.
Since conventional methods (such as strain gauges \cite{JARicodeau1972}, piezoelectric materials \cite{RZLevitin1992}, and capacitance techniques \cite{Gkido1989,MDeorr2008}) - which can be used in pulsed high magnetic fields up to 60 T - are not applicable here due to the severe electromagnetic noise and the extremely fast sweep rate of magnetic field, the FBG method was employed in this study.
The grating part of the optical fiber was fixed to the single crystal of V$_6$O$_{13}$ along the crystallographic $b$-axis by STYCAST1266.
The direction of the magnetic field was also parallel to the $b$-axis of the crystal.

\section{Results}
The temperature dependence of magnetic susceptibility of the single crystal V$_6$O$_{13}$ is shown in Fig. \ref{ChiT_FBG} (a).
The sharp drop indicating the MIT was observed at 153 K
and 165 K in the cooling and heating processes, respectively.
In the lower temperature region, the susceptibility starts decreasing at around 55 K, which is an indication of the antiferromagnetic order.
These behaviors are consistent with those of previous work on V$_6$O$_{13}$ \cite{KKawashima1974,YUeda1976}.

The $\Delta L/L$ of V$_6$O$_{13}$ crystal along the $b$-axis measured by FBG using the same setup for the ultrahigh magnetic field experiment is shown in the Fig. \ref{ChiT_FBG} (b).
A sudden large increase appears with decreasing temperature, as large as 1.4 $\times 10^{-3}$, at $T_{\rm MIT}$.
Previous work on the temperature dependence of the lattice constant of V$_6$O$_{13}$ suggests that $\Delta b / b$ at $T_{\rm MIT}$ is 8.3 $\times 10^{-3}$ \cite{PDDernier1974}.
The observed $\Delta L / L$ in this work is less than the expected $\Delta b / b$ due to the fact that the transmission rate of the lattice change from the sample to FBG is smaller than 100 \%.

Figure \ref{FBG_varT} shows the magnetostriction data measured under ultrahigh magnetic fields at various temperatures.
The magnetostriction as a function of the magnetic field is presented in Fig. \ref{FBG_varT} (a).
No significant change was observed at 20 and 120 K.
At 135 K, only a small negative $\Delta L / L$ appeared near the maximum field.
In contrast, at 145 K and 150 K, $\Delta L / L$ exhibits a sudden large decrease during the field ascending process and a corresponding increase during the descending process.
These changes were accompanied by pronounced hysteresis, which is indicative of a first-order phase transition.
The width of the hysteresis was large, on the order of approximately 50 T, and showed a clear temperature dependence.
The sudden decrease was observed in the ascending process in also the data at 158 K; however, in the field descending process, $\Delta L / L$ remained negative even after the field returned to zero.
Since 158 K lies within the hysteresis region of the MIT in zero field, once the field-induced phase transition occurs, the high-field state can be quenched down to zero field.
In this paper, the onset of the change is defined as the transition point, as indicated by the red and blue triangles in Fig. \ref{FBG_varT} (a).

The $\Delta L / L$ and the magnetic fields as a function of time are shown in Fig. \ref{FBG_varT} (b).
The same data shown in Fig. \ref{FBG_varT} (a) are used.
The change in the data at 120 and 135 K can be more clearly recognized.
The magnetostriction data showing the drastic change (data at 145, 150, and 158 K) accompany the oscillation that occurs after approximately 3 $\mu$s.
This oscillation is caused by a mechanical vibration of the crystal lattice other than the intrinsic magnetostriction as discussed in supplementaly material \cite{Suppl}.

\begin{figure}
\includegraphics[clip,width=0.95\columnwidth]{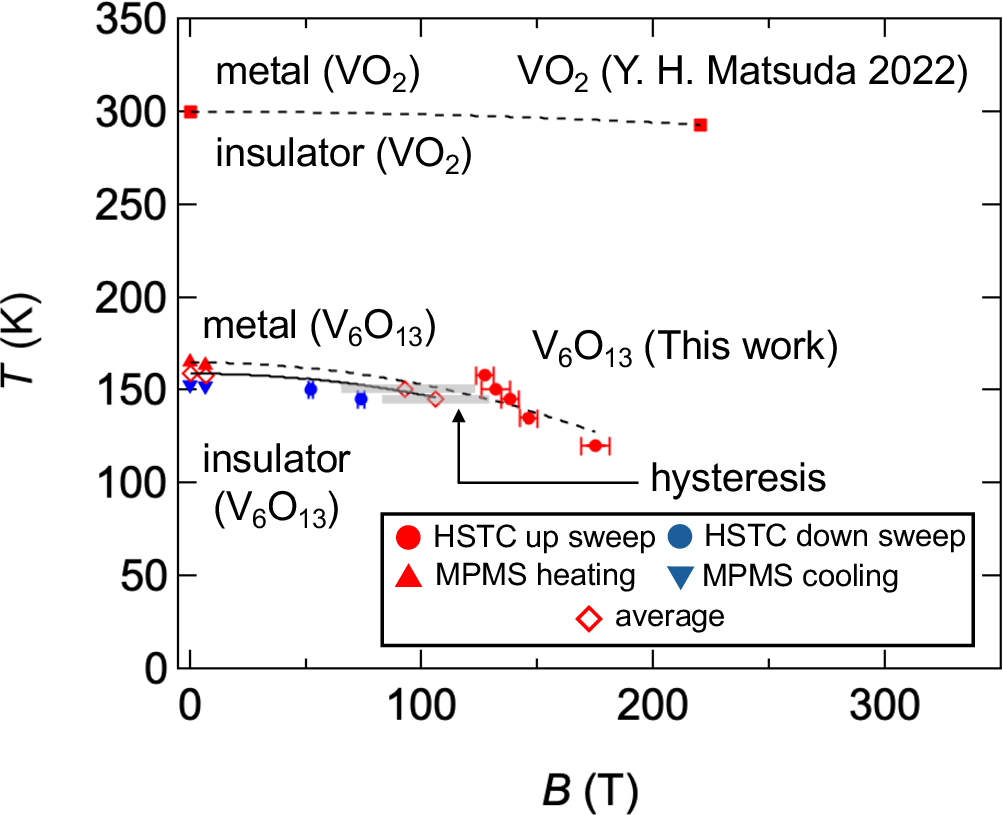}
\caption{\label{PhaDia} (Color online) $B$-$T$ phase diagram of V$_6$O$_{13}$ constructed by this work and that of VO$_2$ constructed by the data imported from Ref.\cite{YHMatsuda2022} (red squares). 
Red and blue circles represent the transition points obtained from the HSTC experiments in the field-ascending and field-descending processes, respectively.
Red and blue triangles denote the transition points determined from the MPMS measurements in the heating and cooling processes, respectively.
The error bars are determined by the standard deviation of the linear fitting of the data and the fluctuation of the magnetic fields. 
Open rhombic symbols indicate transition points obtained by averaging the values between the field ascending and descending processes as well as between the heating and cooling processes.
Broken and solid lines represent the fitting results using a second-order polynomial function for the data in the field ascending process and for the averaged transition points, respectively.}
\end{figure}

The $B$-$T$ phase diagram of V$_6$O$_{13}$ constructed by the ultrahigh magnetic field experimental data is shown in Fig. \ref{PhaDia}.
The phase transition points for VO$_2$ imported from Ref. \cite{YHMatsuda2022} is also shown in Fig. \ref{PhaDia}.

\section {Discussion}

\subsection{Spin singlet in V-V dimer}
The sudden drop of 2.4$\times 10^{-4}$ (emu V-mol$^{-1}$) in the magnetic susceptibility of V$_6$O$_{13}$ at $T_{\rm MIT}$ originates from the formation of a spin-singlet state in the V-V dimer, as shown in Fig. \ref{ChiT_FBG} (a) \cite{TToriyama2014,YShimizu2015}.
The magnitude of this drop is smaller than that observed in typical dimer systems, such as CuGeO$_3$ \cite{MHase1993} and SrCu$_2$(BO$_3$)$_2$ \cite{HKageyama1999}.
This difference arises because, in V$_6$O$_{13}$, the transition occurs from a high-temperature metallic phase to a low-temperature insulating phase with coexistence of residual paramagnetic local spins and spin-singlet pairs, whereas in typical dimer systems the transition is from an insulating paramagnetic state to a fully nonmagnetic state.

\subsection{Field-induced phase transition}
The observed jump in $\Delta L / L$ at $T_{\rm MIT}$ is qualitatively consistent with the temperature dependence of the lattice constant of the $b$-axis reported in a previous X-ray diffraction study on V$_6$O$_{13}$ \cite{PDDernier1974}.
The $b$-axis is not parallel to the V-V dimer axis, and the inter-dimer distance along the $b$-axis increases, resulting in an elongation of the $b$-axis at $T_{\rm MIT}$ (Fig. \ref{image} (b)).
The large $\Delta L / L$ of the order of $10^{-3}$ is comparable to that observed across the MIT associated with V-V dimerization in VO$_2$ \cite{SKitou2024_VO2}.
In this context, the elongation of $b$-axis can be interpreted as a consequence of V-V dimer formation in the MIT of V$_6$O$_{13}$.

Under ultrahigh magnetic fields, a large negative magnetostriction was observed in V$_6$O$_{13}$ as shown in Fig. \ref{FBG_varT}.
Large magnetostriction has also been reported in other systems.
For example, Cr-spinel-based systems exhibit pronounced magnetostriction originating from the magnetic frustration and strong spin-lattice coupling \cite{LRossi2019,MGen2023PNAS}.
In these materials, a first-order magnetic phase transition occurs, leading to the realization of a collinear spin structure \cite{KPenc2004}.
The magnetic properties in V$_6$O$_{13}$ are distinct from those of such systems.
LaCoO$_3$ \cite{AIkeda2020,AIkeda2023} and solid oxygen \cite{AIkeda2025arX} also show the large magnetostriction, arising from spin crossover and from competition between van der Waals and the exchange interactions, respectively.
These mechanisms cannot occur in V$_6$O$_{13}$ because the V$^{4+}$ has $S$ = 1/2 and its bonding is dominated by ionic character (with a finite contribution from covalency).

\begin{figure}
\includegraphics[clip,width=0.95\columnwidth]{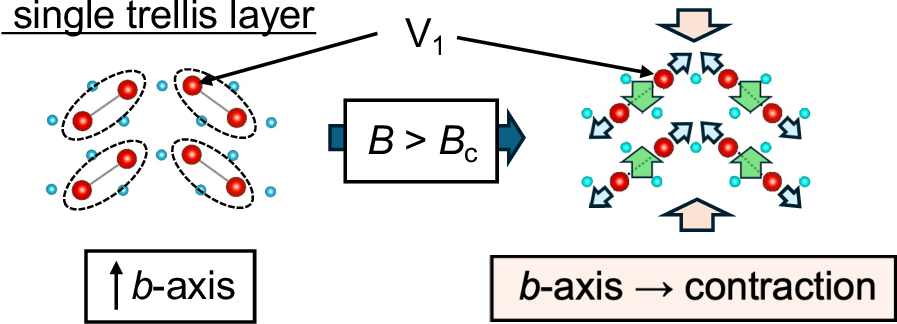}
\caption{\label{contraction} (Color online) Schematic illustration of the expected displacement of V$_1$ ions in the single trellis layer associated with the large negative $\Delta L / L$ induced by ultrahigh magnetic fields along the $b$-axis observed in this work.}
\end{figure}

Another possible scenario is that the large negative magnetostriction arises from the collapse of V-V dimer in the single trellis layer under ultrahigh magnetic fields (Fig. \ref{contraction}).
Indeed, typical dimer systems such as CuGeO$_3$ and SrCu$_2$(BO$_3$)$_2$ show field-induced collapse of spin gap associated with spin singlet dimers \cite{TLorenz1997,HKageyama1999,YHMatsuda2013,TNomura2023}.
In addition, the absolute value of the magnetostriction (1.2 $\times 10^{-3}$) is consistent with the elongation observed in the temperature dependence (1.4 $\times 10^{-3}$).
Furthermore, actually, the field-induced dissociation of V-V dimers has been reported in VO$_2$ \cite{YHMatsuda2020}.
Therefore, this explanation appears to be the most plausible scenario for the large magnetostriction observed in the present work.
In addition, the dome-shaped phase boundary in V$_6$O$_{13}$ (solid and broken black lines in Fig. \ref{PhaDia}) suggests that a field-induced insulator-to-metal transition (IMT) also simultaneously takes place.

\subsection{Chemical catastrophe}
The V-V dimer distance in V$_6$O$_{13}$ (2.984 \AA) is relatively long compared to those in other vanadium oxides - for instance, $\sim$ 2.6 \AA \ in VO$_2$ \cite{TKawakubo1965,SMinomura1964} and $\sim$ 2.7 \AA \ in V$_4$O$_7$) \cite{MMarezio1972}.
Detailed NMR study suggests that the orbital state favors superexchange interactions rather than direct overlap between V ions \cite{YShimizu2015}.
These indicate that the molecular orbital (MO) character of the V-V dimer in V$_6$O$_{13}$ is weaker than that in other vanadium oxides, such as VO$_2$ and V$_4$O$_7$.
As discussed in Ref. \cite{YHMatsuda2020}, the collapse of a MO of a dimer induced by a magnetic field can be interpreted as a non-perturbative magnetic field effect in solids, resembling the phenomenon termed a "chemical catastrophe" \cite{MDate1995}.
This term refers to the collapse of chemical bonding under a very strong magnetic field via the spin Zeeman effect - a process that was once thought to occur only in extremely high fields exceeding 10$^{6}$ T, such as those near neutron stars in cosmic space.
Given that the V-V dimer collapses under magnetic fields, the magnetic field effect in V$_6$O$_{13}$ is expected to be analogous with the physics of the chemical catastrophe, as in the case of VO$_2$ \cite{YHMatsuda2020}.
However, the weakness of MO character and the rich underlying physics - namely, MIT, Mott character, and orbital-assisted Peierls mechanism - makes the phenomenon even more profound.
For a more detailed discussion, further measurements of physical properties over a wider range of temperatures and magnetic fields range are required.

\subsection{Origin of the metal-insulator transition}
In V$_6$O$_{13}$, the charge redistribution, orbital ordering, and V-V dimer formation occur simultaneously across the MIT.
Theoretically, two possible scenarios - band Jahn-Teller and orbital-ordering instability - have been proposed as mechanisms for the MIT \cite{TToriyama2014}.
NMR experiments suggest that the MIT is basically Mott transition, more specifically a site-selective Mott transition \cite{YShimizu2015}.
The formation of V-V dimer spin singlet state within the single trellis layer has been interpreted as an orbital-assisted spin-Peierls transition, which is thought to occur simultaneously with the MIT at $T_{\rm MIT}$ \cite{YShimizu2015,IVLeonov2022}.
Our results indicate that the field-induced IMT coincides with the collapse of the V-V dimer under ultrahigh magnetic fields.
This suggests that the MIT in V$_6$O$_{13}$ arises from the cooperative interplay between electron correlations and structural dimerization, rather than from either mechanism alone.

\subsection{Hysteresis}
The field-induced phase transition in V$_6$O$_{13}$ accompanies the large hysteresis which is greater than 50 T.
This large hysteresis probably comes from the cross-coupling among several types of degrees of freedom, such as spin, charge, orbital, and lattice.
The high sweep rate of the magnetic field ($\sim 10^7$ T s$^{-1}$) is also one of the possible origins of the large hysteresis.
In fact, phase transitions with a structural change induced by ultrahigh magnetic fields always accompany the hysteresis \cite{STakeyama2012,TNomura2014,AIkeda2020,YIshii2023}.
In Ref. \cite{TNomura2014}, it is claimed that the large hysteresis comes from the relaxation time being comparable to the duration time of the pulsed ultrahigh magnetic field \cite{TNomura2014}.
Therefore, in V$_6$O$_{13}$, the relaxation time of the phase transition can also be long enough (comparable to micro seconds).

\subsection{Phase boundary}
Our results indicate that the field-induced phase transition occurs under ultrahigh magnetic fields above 120 T in V$_6$O$_{13}$.
In this section, we discuss the magnetic field effects on V$_6$O$_{13}$ from a thermodynamic perspective, and provide a comparison with VO$_2$, which helps deepen the understanding of field effects on vanadates exhibiting an MIT.
For VO$_2$, it has been shown that the critical magnetic field for filed-induced insulator-metal transition is 220 T at $T$ = 293 K, which is only 7 K lower than $T_{\rm MIT}$ \cite{YHMatsuda2022}.
The phase boundary can be described based on the Clausius-Clapeyron equation as 

\begin{equation}
{\rm d}T/{\rm d}B = -\Delta M/\Delta S
\end{equation}
where $\Delta M$ and $\Delta S$ denote the changes in magnetization and entropy along the phase transition, respectively.
By integrating, we obtain
\begin{equation}
\begin{aligned}
T_{\rm c}(B) = T_{\rm c} (B = 0) - (\Delta \chi \cdot B^2) / (2\Delta S) 
\end{aligned}
\end{equation}
under the assumption that $\Delta M = \Delta \chi \cdot B$.
Fitting the experimental phase boundaries using a simple second-order polynomial, ($T_{\rm c} = T_{\rm c}(B = 0) - A \cdot B^2$), as shown in Fig. \ref{PhaDia}, (where $A$ = ($\Delta \chi$) / (2$\Delta S$)) yields $A$ = 1.2 $\times 10^{-3}$ (K T$^{-2}$) for V$_6$O$_{13}$ (for both the solid and broken lines in Fig. \ref{PhaDia}) and 1.4 $\times 10^{-4}$ (K T$^{-2}$) for VO$_2$. We calculated the $A$ using the reported values of $\Delta \chi$ and  $\Delta S$ measured in the absence of magnetic fields (Table \ref{tab01} ) \cite{ZHiroi2015,Howing2003,GVChandrashekhar1973,Uchimura2020}.

\begin{table}[htbp]
\centering
\caption{$\Delta \chi$ (emu V-mol$^{-1}$) and $\Delta S$ (J K$^{-1}$ V-mol$^{-1}$) across metal-insulator transitions of V$_6$O$_{13}$ and VO$_2$, calculated ($A$ (cal.)) and experimental ($A$ (exp.)) values of $A$ (=$\Delta \chi / 2 \Delta S$ (K T$^{-2}$)) of V$_6$O$_{13}$ and VO$_2$.}
\normalsize
\begin{tabular}{cccccc} \hline \hline
 & $\Delta \chi$ \quad & $\Delta S$ \quad & $A$ (cal.) \quad & $A$ (exp.) \quad \\ \hline
V$_6$O$_{13}$ \quad & 2.4 $\times 10^{-4}$ \quad & 2.16 \quad & 5.6 $\times 10^{-4}$ \quad & 1.2 $\times 10^{-3}$ \quad  \\
VO$_2$ \quad & 5.5 $\times 10^{-4}$ \quad & 13.77 \quad & 2.0 $\times 10^{-4}$ \quad & 1.4 $\times 10^{-4}$ \quad \\ \hline \hline
\end{tabular}
\label{tab01}
\end{table}
\noindent
Although this calculation is based on the simplified assumption, the calculated values of $A$ are semiquantitatively in good agreement with the experimentally determined values.
The robustness of the insulating phases against magnetic fields can be understood as a consequence of the small $\Delta \chi$ and the large $\Delta S$.
The small $\Delta \chi$ is attributed to the fact that the high-temperature phases of both V$_6$O$_{13}$ and VO$_2$ are metallic and therefore exhibit small magnetization, as discussed earlier.
The large $\Delta S$ originates from the simultaneous occurrence of the structural transition and spin-singlet formation along the MIT in both V$_6$O$_{13}$ and VO$_2$.
The difference between V$_6$O$_{13}$ and VO$_2$ can be attributed to the unusually large phonon and electronic entropy contributions in MIT of VO$_2$ \cite{JDBudai2014,JParas2020}.

\subsection{Orbital-assisted spin-Peierls state}
The orbital-assisted spin-Peierls mechanism has been discussed for V-V dimers in V$_6$O$_{13}$ \cite{TToriyama2014,YShimizu2015}.
This mechanism has been also proposed for spinel based materials exhibiting MIT such as MgTi$_2$O$_4$ and CuIr$_2$S$_4$ \cite{DIKhomski2005}.
However, since a ultrahigh magnetic field study of MgTi$_2$O$_4$ and CuIr$_2$S$_4$ has never been performed so far, it is difficult to discuss perspectives of field effects on the orbital-assisted Peierls states.
Actually, a theory for a pure spin-Peierls system \cite{MCCross1979-2} and an experiment in the case of CuGeO$_3$ \cite{TLorenz1997} revealed that the properties of $B$-$T$ phase diagram and appearance of the incommensurate phase.
However, in V$_6$O$_{13}$, the high-temperature phase is metallic and the V-V chain forms a zigzag structure; thus, the situation is distinct from that of a pure spin-Peierls system.
Nevertheless, discussion based on the similarities with spin-Peierls theory remain important. In particular, the possibility of a new type of incommensurate phase is intriguing in the context of the field-induced insulator-to-metal transition.
Even though further experimental data and modifications of the spin-Peierls framework are required for a more detailed understanding, this work can trigger a study on magnetic field effects on orbital-assisted Peierls state.

\section{Conclusion}
We have measured the magnetostriction of V$_6$O$_{13}$ under ultrahigh magnetic fields of up to 186 T.
The large negative $\Delta L / L$ along the $b$-axis was observed above 120 T below $T_{\rm MIT}$.
The observed $\Delta L / L$ can be interpreted as a consequence of the collapse of the V-V dimer under ultrahigh magnetic fields accompanied by the insulator-to-metal transition.
The pronounced hysteresis is probably because of the cross-coupling among the several types of degrees of freedom.
The $B$-$T$ phase diagram was discussed based on the thermodynamics using Clausius-Clapeyron equation and the comparison with the case of VO$_2$.
This study provides a new experimental demonstration of the non-perturbative magnetic field effect on electronic dimer in solids and indicates its utility in understanding the physical properties of materials.

Another important point to be studied is the direct observation of insulator-to-metal induced by ultrahigh magnetic fields in V$_6$O$_{13}$.
However, the standard technique for resistance measurement is not suitable due to the huge noise coming from the explosion of single-turn coil.
Recently, radio-frequency (RF) reflection-type impedance measurement technique under ultrahigh magnetic fields above 100 T has recently been developed \cite{SPeng2025,TShitaokoshi2023}.
We can investigate the insulator-to-metal transition directly using the new RF technique.
Furthermore, the higher magnetic field experiment using the electro-magnetic flux compression system up to 1200 T \cite{DNakamura2018} at a lower temperature region is also important and interesting for the quest for possible new quantum phases.
 \\

\begin{acknowledgments}
We thank H. Oike, M. Gen, N. Matsuyama for fruitful discussions, R. Ishii for help of the single crystal preparation, J. Yamaura for help of the XRD experiment. 
The schematic image of crystal structures were drawn by the VESTA software  \cite{KMomma2011}.
This research was supported by JSPS KAKENHI Grants Numbers 23H01117, JP23H04859, JP23H04860, JP23H04861, 24K17003, and JP25H02118.
\end{acknowledgments}

\bibliography{V6O13}

\clearpage
\onecolumngrid
\appendix

\section{Supplementary material}
\maketitle

In this supplemental material, we will introduce the information about the sample preparation, detailed FBG experimental setup information, and magnetic field dependence of magnetic susceptibility of V$_6$O$_{13}$.

\subsection{sample preparation}
The single crystalline samples of V$_6$O$_{13}$ were prepared by chemical vapor transport method using TeCl$_4$.
Temperatures of high temperature and low temperature regions are 600$^{\cdot}$C and 550$^{\cdot}$C, respectively.
It was kept for 1 week, after that we obtained needle like single crystals in the low temperature side.
Basically, we followed the reported way in Ref \cite{KNagasawa1969}.

\subsection{Details of Fiber Bragg Grating experiment and discussion for oscillation}

\begin{figure*}[h]
\includegraphics[clip,width=0.5\columnwidth]{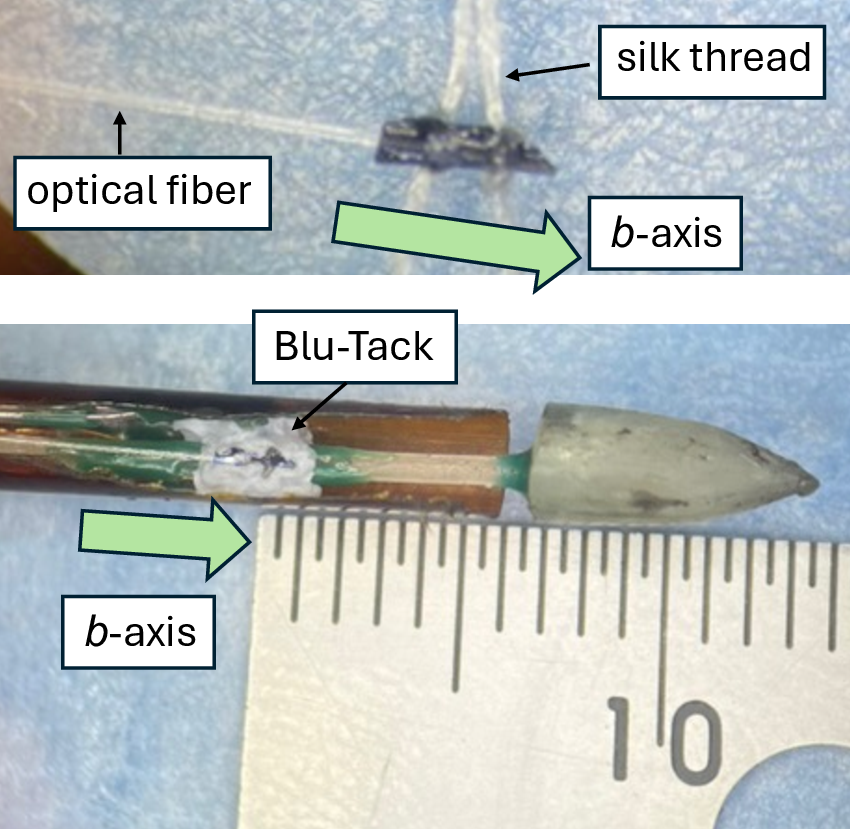}
\caption{\label{FBGpicture} (Color online)  Pictures of the FBG attached to V$_6$O$_{13}$ single crystal using STYCAST1266 as shown in the top panel. The silk threads were used to hold the FBG and the crystal in place while the STYCAST1266 cures. The bottom panel shows the probe, where the single crystal was fixed to the probe using Blu-Tack (a light blue adhesive compound).}
\end{figure*}

Figure \ref{FBGpicture} is the picture of the V$_6$O$_{13}$ single crystal sample fixed to the FBG by STYCAST1266.
Thin white wires are silk wires to hold the FBG and sample during waiting for drying the STYCAST1266.

\begin{figure*}
\includegraphics[clip,width=0.6\columnwidth]{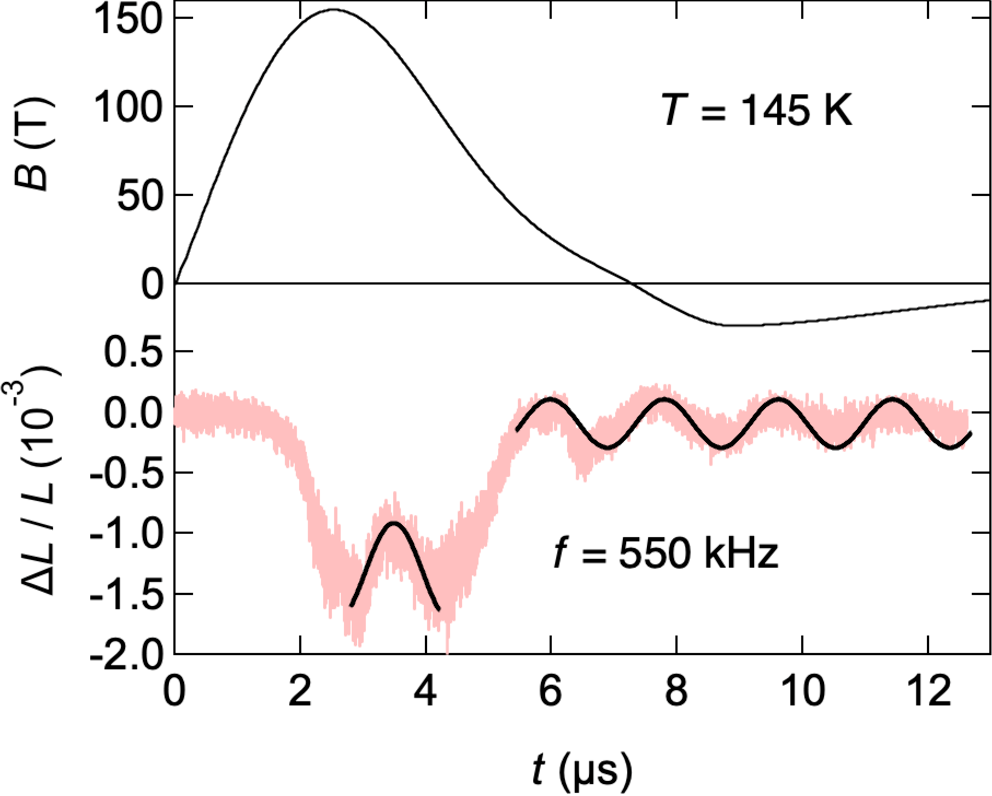}
\caption{\label{FBGOsc} (Color online) FBG data and magnetic field as a function of time. Black lines are the simple sine function. The oscillation can be reproduced by the simple sine function with $f \sim 550$ kHz.}
\end{figure*}

Figure \ref{FBGOsc} shows the magnetic field and $\Delta L / L$ as a function of time.
The $\Delta L / L$ shows the oscillation with the frequency of around 600 ($\pm$ 100) kHz.
We think that the oscillation is the vibration due to the mechanical resonance.

Generally, the resonance frequency is given by

\begin{equation}
f_n = \frac{n}{2L}c_b
\end{equation}

Here, we used the sound velocity of $c_b \sim 5 \times 10^3$ m/s which is the typical value for the inorganic material, and the sample length $L$ = 2$\times 10^{-3}$ m.
This yields an estimate of $f_1 \sim10^6$ Hz, which is roughly consistent with the frequency of the observed oscillation.
The observed frequency is slightly lower than the estimation probably due to that the sample and FBG is covered by STYCAST1266.
Such compounds, including epoxy-based materials, generally exhibit lower sound velocities ($\sim 2\times10^3 - 3\times10^3$ m/s).
Therefore, the observed oscillation can be interpreted as a standing wave arising from mechanical resonance.
A similar discussion has been reported in previous studies on magnetostriction measurements under ultrahigh magnetic fields \cite{RSchonemann2021,MGen2023PNAS}.

\end{document}